\documentclass[aps,twocolumn,showpacs,floatfix]{revtex4}
\usepackage{amsmath,amsfonts,graphicx,psfrag,psfig,bm}
\tighten      
\begin{document}      
%
\title{%
\hfill{\normalsize\vbox{%
\hbox{\rm UH-511-1004-02}
\hbox{\rm SLAC-PUB-9283} 
\hbox{\rm FSU-HEP-020709}}} \\

\vspace{0.3cm}
{Relating bottom quark mass in $\overline{\rm DR}$ and 
$\overline{\rm MS}$
regularization schemes} }     
\author{Howard Baer$^1$, Javier Ferrandis$^2$, Kirill Melnikov$^3$ and Xerxes Tata$^2$}      
\address{      
$^1$Department of Physics,
Florida State University,
Tallahassee, FL 32306, U.S.A.}
\address{      
$^2$Department of Physics \& Astronomy      
University of Hawaii,      
Honolulu, HI 96822, U.S.A.}      
\address{$^3$Stanford Linear Accelerator Centre, Stanford University,
Stanford, CA 94309, U.S.A.}         
%
\begin{abstract}    
The value of the bottom quark mass at $Q=M_Z$ in the    
$\overline{\rm DR}$ scheme is an important input for the analysis of    
supersymmetric models with a large value of $\tan\beta$. Conventionally,    
however, the running bottom quark mass extracted from experimental data is    
quoted in the $\overline{\rm MS}$ scheme at the scale $Q=m_b$.     
We describe a two loop procedure for the conversion of    
the bottom quark mass from $\overline{\rm MS}$ to     
$\overline{\rm DR}$ scheme. The Particle Data Group value    
$m_b^{\overline{\rm MS}}(m_b^{\overline{\rm MS}})= 4.2\pm 0.2$~GeV corresponds to a    range of 2.65--3.03~GeV for $m_b^{\overline{\rm DR}}(M_Z)$.    
\end{abstract}      
\maketitle     
\medskip      
\newpage      

\section{Introduction}      
      
Supersymmetry (SUSY) is one of the best-studied extensions of the    
Standard Model (SM)~\cite{rev}. Despite the lack of direct evidence, many    
theorists consider the idea promising enough to examine even quantum    
corrections in SUSY theories: these include corrections to SM quantities    
due to sparticle loops~\cite{pierce}, as well as corrections to processes    
involving the production and decays of supersymmetric    
particles~\cite{zerwas}. Quantum corrections in SUSY theories are also    
important for obtaining the implications of top-down models, where    
simple assumptions about symmetries of physics at the high scale result    
in a highly predictive scenario. The best known example of these is the    
correction~\cite{carena} to the relation between the down type mass and    
the corresponding superpotential Yukawa coupling, which is crucial to    
include when analysing SUSY models for large values of the parameter    
$\tan\beta$.    
      
For large $\tan\beta$ values, third generation sparticle and Higgs boson  
masses as well as sparticle decay patterns, and thus collider signals  
for SUSY, are sensitive to the value of the bottom (and to a lesser  
extent tau) Yukawa coupling~\cite{large}. Knowledge  
of Yukawa couplings is also necessary for an analysis of the interesting  
possibility that these are unified at a high scale. Yukawa coupling  
unification may, for instance, be realized  
in the minimal $SO(10)$ SUSY model provided $\tan\beta$ is large~\cite{soten}.  
The interpretation of super-Kamiokande atmospheric neutrino  
data~\cite{superk} in terms of neutrino oscillations originating in a  
non-trivial flavour structure of the neutrino mass matrix has led to  
considerable recent interest~\cite{us,raby} in such a scenario.

To include radiative corrections in the phenomenological analysis of  
supersymmetric models, the $\overline{\rm DR}$ regularization by dimensional  
reduction~\cite{Siegel:1979wq}, a modification of the conventional  
dimensional regularization is commonly used as a candidate for an  
invariant regularization of supersymmetric theories. Since the bottom  
quark Yukawa coupling is determined from the running bottom mass  
parameter at low energy, a careful determination of $m_b(M_Z)$ in the  
$\overline{\rm DR}$ scheme is necessary.  
      
On the other hand, it became a standard practice to use experimental data to
extract the value of the running bottom mass parameter $m_b(m_b)$ in the
$\overline{\rm MS}$ renormalization scheme. The most recent results based
on NNLO QCD  analyses of the $\Upsilon$ sum rules can be found in 
~\cite{mass} (see also the summary in ~\cite{Groom:in} which, however, 
also includes older (and by now obsolete) results in its world average). 
It is useful to
explicitly examine what the allowed range of $m_b^{\overline{\rm MS}}(m_b)$
\cite{note} translates into when expressed in terms of
$m_b^{\overline{\rm DR}}(M_Z)$. While an immediate motivation to study this
is because the allowed range of $m_b^{\overline{\rm DR}}(M_Z)$ yields an
estimate of the ``error'' within which Yukawa couplings in SUSY models
with Yukawa coupling unification should be required to unify, our result
for $m_b^{\overline{\rm DR}}(M_Z)$ would be of relevance for any SUSY
analysis with large $\tan\beta$.

Many authors~\cite{aronson} start with the value of the bottom
quark mass $m_b^{\rm pole}$ to extract $m_b^{\overline{\rm DR}}(M_Z)$,
either by directly relating the pole mass to the running mass at the
high scale (as described in Sec.~\ref{running}), 
or by first using the pole mass
to derive $m_b^{\overline{\rm DR}}$ at the scale of the
bottom quark mass and then evolving it to $M_Z$ using renormalization
group equations.  Since it became clear in recent years that the pole
mass of the quark suffers from infra-red sensitivity which precludes its
accurate determination from physical observables, this approach leads to
significant ambiguities in the resulting value of $m_b^{\overline{\rm
DR}}(M_Z)$.  In this paper, we suggest a better approach to the
determination of $m_b^{\overline{\rm DR}}(M_Z)$, which by-passes the use
of the $b$-quark pole mass.
The idea is to employ  recent accurate determinations 
of  $m_b(m_b)$ in the $\overline{\rm MS}$ scheme \cite{mass}, 
use the renormalization 
group running up to the scale $\mu=M_Z$ and use a relation between 
the $\overline{\rm MS}$ and $\overline{\rm DR}$ masses at that 
scale to obtain $m_b^{\overline{\rm DR}}(M_Z)$. We will show 
that the relation that connects  the masses in the two schemes 
at the same scale $\mu$ exhibits excellent convergence properties 
thereby making the suggested  procedure very accurate and 
practically insensitive to higher order corrections.

\section{$\overline {\rm MS}$ 
and $\overline {\rm DR}$  mass formulas}\label{running}    
Our final goal will be a relation between the $\overline {\rm MS}$ 
and $\overline {\rm DR}$ masses. However, since in the literature 
the two masses are usually related to the pole mass,       
we begin by collecting formulae that relate the pole
mass of the bottom quark with its running mass (at any scale) in both
the $\overline{\rm MS}$ and $\overline{\rm DR}$ schemes. 

Currently, the relation between the $\overline{\rm MS}$
 and pole quark mass is known
through  order ${\cal O}(\alpha_s^3)$  
~\cite{Tarrach:1980up,Gray:1990yh,Fleischer:1998dw,Chetyrkin:1999qi,Chetyrkin:2000yt,Melnikov:2000qh}. In this paper we will only use the two loop 
relation between them:
\begin{eqnarray}
&& \frac{m_b^{\overline{\rm MS}} (\mu)}{m_b^{\rm pole}} = 1
+a_s(\mu)
\left (L -\frac{4}{3}   \right )
+a_s^2(\mu)
\left( -\frac{11}{24}L^2 
\right.  \nonumber \\
&& \left. 
+ \frac{197}{72}L
-\frac{187}{32}+\frac{\zeta_3}{6} -\frac{\pi^2}{9}\ln 2 -\frac{\pi^2}{9}
   -\Delta  \right ),
\label{chetyrkininv}      
\end{eqnarray}
where $a_s(\mu) = \alpha_s(\mu)/\pi$ 
is the $\overline {\rm MS}$ coupling constant
in the theory with five quark flavors, $L=\log(m_b^2(\mu)/\mu^2)$,
and $\Delta$ is a correction due to light quark mass effects
(arising from the two-loop diagram with a second
fermion-loop), given by~\cite{Gray:1990yh}, 
\begin{eqnarray}      
\Delta  &=& \sum_{i=u,d,s,c}      
\delta  \left( \frac{m_i}{m_b} \right)\;, \\      
\delta (x) &=& 
\frac{\pi^2}{6}x(1+x^2)-x^2 + {\cal O}(x^4).
\label{deltams}    
\end{eqnarray}      
In the argument of $\Delta$ the ratio 
of quark masses of the light quarks
to the bottom quark pole mass appear.  For $0\le x \le 1$ the exact 
result is approximated by the three terms given in 
Eq.~(\ref{deltams}) within an accuracy of $1$\%.

The relation between the pole mass and the quark running masses      
in the $\overline{\rm DR}$ renormalization scheme  is known 
through ${\cal O}(\alpha_s^2)$~\cite{Avdeev:1997sz}. 
The result reads:      
\begin{eqnarray}      
&& \frac{m_b^{\rm pole}}{m_b^{\overline{\rm DR}} (\mu)} =  
 1 +      
\bar a_s(\mu) \left ( 
\frac{5}{3} -L_1\right )     
+ \bar a_s^2(\mu)
\left (  \frac{35}{24}L_1^2
\right. \nonumber \\
&& \left.
-\frac{395}{72}L_1
+\frac{2159}{288}
-\frac{\zeta_3}{6}+\frac{\pi^2}{9} + \frac{\pi^2}{9}\ln2
+\Delta
\right ),
\label{avdeev}      
\end{eqnarray}      
where $\bar a_s(\mu) = \bar \alpha_s(\mu)/\pi$ 
is the strong coupling constant in the $\overline {\rm DR}$
scheme and $L_1 = \ln(m_b^{\rm DR}(\mu)^2/\mu^2)$. 

It is possible to invert 
Eq.~(\ref{chetyrkininv}), expressing $m_b^{\rm pole}$ in terms of
$m_b^{\overline{\rm MS}}(m_b)$, and then use this together with      
Eq.~(\ref{avdeev}) choosing $\mu=M_Z$ 
to obtain $m_b^{\overline{\rm DR}}(M_Z)$. This 
is unsatisfactory for two reasons.  First of all, 
the relation between $m_b^{\rm pole}$  and 
$m_b^{\overline{\rm DR}}(M_Z)$ involves not so small      
logarithm ($C_F \bar a_s(M_z) \log(M_Z^2/m^2) \sim 0.3$) which leads 
to uncomfortably large corrections.  Second, the 
perturbative corrections to the  relation between $m_b^{\rm pole}$ and 
$m_b^{\overline{\rm MS}}(m_b)$ are  large. For example, including  
the ${\cal O}(\alpha_s^3)$ term in Eq.~(\ref{chetyrkininv})
leads to a correction comparable to the 
contribution of the ${\cal O}(\alpha_s^2)$ term. By the same token, 
when inverting Eq.~(\ref{chetyrkininv}) to determine the value of 
$m_b^{\rm pole}$ out of $m_b^{\overline{\rm MS}}$, it makes a significant 
numerical difference if one expands the inverse equation in powers
of $a_s(\mu)$ or just keeps the r.h.s. of  Eq.~(\ref{chetyrkininv})
in the denominator. These problems show  that, as a consequence 
of large perturbative corrections,   the 
final {\it numerical} results for $m_b^{\overline{\rm DR}}(M_Z)$ 
when obtained from Eq.~(\ref{chetyrkininv}), are not stable 
against higher orders perturbative corrections.
Analyses ~\cite{aronson}  that use the pole mass as an input for extracting 
the short-distance mass, and via this, quark Yukawa couplings, 
will reflect this ambiguity. 
In what follows we suggest a procedure 
wherein the problems just mentioned are considerably ameliorated.
      
\section{$\overline{\rm DR}$ mass from the 
$\overline{\rm MS}$ bottom mass}      
      
The first of the above problems can be  by-passed because      
an analytic solution of the renormalization      
group equation for the running mass in the $\overline{\rm MS}$ scheme is      
known. This was originally obtained       
at three loops ~\cite{Tarasov:au} and  recently the four loop term      
has also been computed ~\cite{Chetyrkin:1997dh}.      
For the case of five active quark flavors, 
and up to NNLO renormalization
group evolution, this takes the form      
\begin{eqnarray}      
m_b^{\overline{\rm MS}} (\mu) = &&       
\hat {m}_{\rm MS} \left(\frac{23a_s(\mu)}{6} \right)^{12/23}      
\left(      
1 + \frac{3731}{3174}~a_s(\mu) 
\right.  \nonumber \\
&& \left. 
+ 1.500706~a_s^2(\mu)\right)\;,      
\label{mbottomrunms}      
\end{eqnarray}      
where the integration constant $\hat {m}_{\rm MS}$      
is the renormalization      
group invariant bottom mass. We do not need to know $\hat {m}_{\rm MS}$ because,      
if we denote the right hand side as $\hat {m}_{\rm MS} F_b(\mu)$, the running      
mass at scale $M_Z$ can be calculated      
from a given running mass at scale $m_b(m_b)$      
using the expression,      
\begin{equation}      
m_b^{\overline{\rm MS}} (M_Z) = m_b^{\overline{\rm MS}}(m_b) 
\frac{F_b(M_Z)}{F_b(m_b)}      
\label{mbmzmbmb}      
\end{equation}      
Although Eq.~(\ref{mbottomrunms}) has been written 
in the $\overline{\rm MS}$  scheme,      
we observe that the leading term, determined entirely by the 
first coefficient of the $\beta$-function $\beta_0$ and 
the first coefficient of the quark mass anomalous dimension $\gamma_0$, 
is scheme-independent. At $Q=M_Z$ the      
scheme-dependent subleading terms contribute $\sim 8$\%. Most of this
comes from the next to leading $\alpha_s/\pi$ term while the three  
loop term  contributes just 0.2\%.
The coefficient of this two loop term can be readily evaluated in the 
$\overline{\rm DR}$ scheme
using,
\begin{equation}      
\gamma_1^{\overline{\rm DR}} = \gamma_1 -\frac{\beta_0}{3}
-\frac{\gamma_0}{4}\;,
\label{gamma1dr}      
\end{equation}      
where  $\gamma_1$ is the coefficient of the ${\cal O}(\alpha_s^2)$ 
term in  the quark mass 
anomalous dimension. At two loops,    
we obtain for the running bottom mass in the $\overline{\rm DR}$ 
scheme, 
\begin{equation}      
m_b^{\overline{\rm DR}} (\mu) \equiv      
\hat {m}_{\rm DR} \left(\frac{23 \bar a_s(\mu)}{6} \right)      
^{12/23}      
\left(      
1 + \frac{753}{1058}~\bar a_s(\mu)       
\right)\;.     
\label{mbottomrundr}      
\end{equation}      

To compute the bottom running mass at $m_b$ we need      
the value of $\alpha_s(m_b)$ that     
corresponds to the experimental measurement,      
$\alpha_s(M_Z)$.      
To this end we will use the three loop      
analytical formula for $\alpha_s$ ~\cite{Tarasov:au} in  the
$\overline{\rm MS}$ scheme, that is the     
solution of the corresponding Callan-Symanzyk equation.      
In the case of five active quark flavours the formula reduces      
to,      
\begin{eqnarray}      
&& \alpha_s(\mu) = \frac{12\pi}{23 t}      
\left [      
1 - \frac{348}{529}\frac{\ln (t)}{t}      
\right. \nonumber \\
&& \left. + \left( \frac{348}{529} \right)^2 \frac{1}{t^2}      
\left( \left( \ln(t) -\frac{1}{2}\right)^2      
- \frac{78073}{242208} \right)      
\right ],      
\label{alfasrun}      
\end{eqnarray}      
where $t=\ln \left(\mu^2/\Lambda^2 \right)$. In practice, it is      
convenient to use the value of $\alpha_s(M_Z)$ to first determine      
$\Lambda$, and then use Eq.~(\ref{alfasrun}) again for the calculation      
of $\alpha_s(m_b)$ which is inserted in Eq.~(\ref{mbmzmbmb}) before the      
iteration.  The four loop contributions to Eq.~(\ref{alfasrun}) have      
also been calculated ~\cite{Vermaseren:1997fq} and found to be small. It      
was observed that the previous three loop analytical solution for      
$\alpha_s(\mu)$ gives a very good aproximation to the numerical solution      
of the four loop RGE ~\cite{Chetyrkin:1997sg}.        
Finally, we note that to compute      
$\alpha_s(\mu)$ in the $\overline{\rm DR}$ scheme we convert      
$\alpha_s(\mu)_{\overline{\rm MS}}$ using ~\cite{Antoniadis:1982vr},      
\begin{equation}      
\frac{1}{\bar a_s(\mu)} =      
\frac{1}{a_s(\mu)} -      
\frac{1}{4}\;.      
\label{drmsconv}      
\end{equation}      
By eliminating the pole mass       
from the relations between the pole and running masses      
presented in the previous section,  we can derive      
a two loop relation between the $\overline{\rm DR}$ 
and the $\overline{\rm MS}$      
bottom masses at the {\it same} scale.
>From (\ref{avdeev}) and
(\ref{chetyrkininv}), using Eq.~(\ref{drmsconv}) to convert the 
$\overline {\rm DR}$ coupling constant to the 
$\overline {\rm MS}$ coupling constant,
we obtain:      
\begin{equation}      
 m_b^{\overline{\rm DR}} (\mu) = m_b^{\overline{\rm MS}} (\mu)      
\left( 1 - \frac{1}{3}a_s(\mu)    
- \frac{29}{72} a_s(\mu)^2
+{\cal O}(a_s^3)
\right).      
\label{mbconvmsdr}      
\end{equation}      
This expression also appears, but is never used, in Ref.~\cite{Avdeev:1997sz}.

We were originally motivated to consider the relation between
$\overline{\rm MS}$--$\overline{\rm DR}$ masses because it was clear that the
leading logarithm corrections are absent because leading
contributions to the QCD beta function and quark mass anomalous dimension
are scheme-independent.
The absence of {\it all} $\log \mu/m_b$ terms follows from 
the fact that both $\overline{\rm MS}$ and $\overline{\rm DR}$
are mass-independent  renormalization schemes and that
$(\gamma^{\overline{{\rm DR}}}-\gamma^{\overline{{\rm MS}}})/
\beta^{\overline{{\rm MS}}}$ is regular
around $\alpha_s=0$~\cite{note2}. The 
renormalization group equations then ensure that the relation 
between the two masses at the same scale can only depend on the 
coupling constant $\alpha_s(\mu)$. 

The perturbative series in  Eq.~(\ref{mbconvmsdr}) exhibits 
excellent convergence of the perturbative expansion. 
This is to be expected,  since the $\overline{\rm MS}$ 
and the $\overline{\rm DR}$ mass are both
truly short-distance quantities. The infrared 
renormalon problem,  which is the reason for  
large perturbative corrections in the  relations between 
the pole mass of the quark and any of the short-distance
masses, is not relevant there. Note that the  cancellation 
of large corrections happens order by order in perturbation theory 
and for this reason one cannot improve on 
the obtained relation by including {\it e.g.} the three loop term from 
the relation between the pole and the $\overline{\rm MS}$ mass without
including a similar term from the relation between 
the pole and the $\overline{\rm DR}$ mass.

These considerations suggest the following two step procedure for
extracting $m_b^{\overline{\rm DR}}(M_Z)$, starting from
$m_b^{\overline{\rm MS}}(m_b)$.
\begin{itemize}      
\item First, evolve 
$m_b^{\overline{\rm MS}}(m_b)$ to $Q=M_Z$ using      
(\ref{mbottomrunms}) which is valid within the 
$\overline{\rm MS}$ scheme.      
\item Use Eq.~(\ref{mbconvmsdr}) to convert $m_b^{\overline{\rm
MS}}(M_Z)$ thus obtained to $m_b^{\overline{\rm DR}}(M_Z)$.
\end{itemize}
In principle, we could have by-passed the first step by directly using the
determination ~\cite{delphi} of $m_b^{\overline{\rm MS}}(M_z)$ from three jet
$b$-quark production at LEP, but the associated errors are so large that
we cannot use it for our purposes.

\begin{figure}[bht]    
\begin{minipage}{16.cm}
\psfig{figure=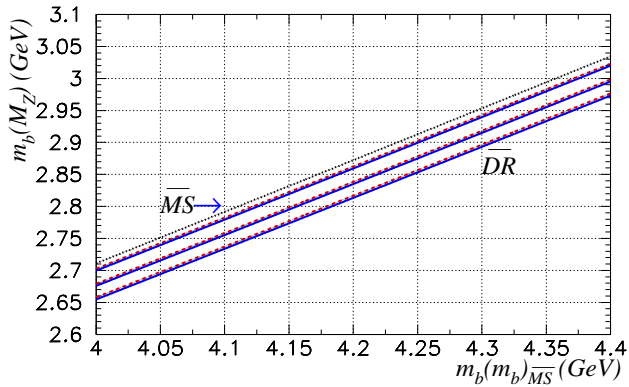,width=90mm}
\end{minipage}
\caption{\it     
$m_b(M_Z)$ as a function of $m_b^{\overline{\rm MS}} (m_b)$.    
The solid lines show the results for $m_b^{\overline{\rm DR}}(M_Z)$ 
using the first procedure described in the    
text, where we first run $m_b^{\overline{\rm MS}}$ to $M_Z$, and then    
convert to the $\overline{\rm DR}$ mass at the same scale. The lines    
correspond from top to bottom to values of     
$\alpha_s(M_Z)_{\overline{\rm MS}}=0.1152,0.1172,0.1192$. The dashed lines    
show the corresponding results when we first convert to    
the $\overline{DR}$ scheme, and then evolve to $Q=M_Z$. Finally, the
dotted line shows the value of $m_b^{\overline{\rm MS}} (M_Z)$ for 
$\alpha_s(M_Z)_{\overline{\rm MS}}=0.1172$. }    
\label{fig:botdrms}    
\end{figure}

The  value of $m_b^{\overline{\rm DR}}(M_Z)$ obtained using this two-step
procedure is shown as a function of $m_b^{\overline{\rm MS}}(m_b)$ by the
central solid line in Fig.~\ref{fig:botdrms}. This line is obtained,
using Eq.~(\ref{mbconvmsdr}), for $\alpha_s
(M_Z)_{\overline{\rm MS}}=0.1172$. The upper (lower) solid lines correspond
to a choice for $\alpha_s$ that is smaller (larger) than its central
value by 0.002. For comparison, we also show by the dotted line the
value of $m_b^{\overline{\rm MS}}(M_Z)$ using $\alpha_s
(M_Z)_{\overline{\rm MS}}=0.1172$.

%
%
An alternative approach would be to derive
$m_b^{\overline{\rm DR}}(m_b)$ using Eq.~(\ref{mbconvmsdr}) and then 
evolve it to $Q=M_Z$
using Eq.~(\ref{mbottomrundr}). If we do so, we obtain the dashed lines in
the figure. The relative difference between these and the solid lines is
similar to the size of the three loop term  in
Eq.~(\ref{mbottomrunms}). The agreement between the dashed and solid lines
provides additional evidence for the reliability of our procedure.

\section{Summary}       
In summary, we have suggested a procedure to convert the value of      
$m_b^{\overline{\rm MS}}(m_b)$ that is extracted from experimental      
data to $m_b^{\overline{\rm DR}}(M_Z)$. This value      
serves as an important input for SUSY analyses with large $\tan\beta$.     
    
Our main results are summarized in Fig.~\ref{fig:botdrms}. 
Most of the recent analyses indicate \cite{mass} 
that  $m_b^{\overline{\rm MS}}(m_b)= 4.2\pm 0.1~{\rm GeV}$. This translates
to $m_b^{\overline{\rm DR}}(M_Z)=2.83\pm 0.11$~GeV. 
If instead we take the Particle Data Group range~~\cite{Groom:in}
$4.0-4.4$~GeV for $m_b^{\overline{\rm MS}}(m_b)$, the range of
$m_b^{\overline{\rm DR}}(M_Z)$ extends 
from $2.65-3.03$~GeV.

{\bf Acknowledgments} 

We thank M.~Kalmykov for 
instructive correspondence, and C.~Bal\'azs,
A.~Belyaev and T.~Blazek for helpful discussion.  This research was
supported in part by the U.S. Department of Energy under grants
DE-FG03-94ER40833, DE-FG02-97ER41022 and DE-AC03-76-SF00515.
      
%

\begin{references}      
%
\bibitem{rev} For recent reviews, see {\it e.g.} S.~Martin,      
in {\it Perspectives on Supersymmetry}, edited by G.~Kane (World Scientific),      
[hep-ph/9709356]; M.~Drees, [hep-ph/9611409];      
J.~Bagger, [hep-ph/9604232];      
X.~Tata, {\it Proc.~IX J.~Swieca Summer School,}      
J.~Barata, A.~Malbousson and S.~Novaes, Eds., [hep-ph/9706307];      
S.~Dawson, {Proc. TASI 97}, J.~Bagger, Ed., [hep-ph/9712464].      
%
\bibitem{pierce} Some examples include, D.~Pierce {\it et al.}    
Nucl. Phys. {\bf B491}, 3 (1997); {E-Proc. RADCOR 2000, H.~Haber,  
Editor}, http://www.slac.stanford.edu/econf/C000911/;  
 {\em Radiative Corrections:    
Application of Quantum Field Theory to Phenomenology, Proc. RADCOR 98},    
J. Sola, Editor (World Scientific, 1999); {\em Quantum Effects in the    
Minimal Supersymmetric Standard Model, Proc. Int. Workshop on the MSSM},    
Barcelona, Spain, 1997, J.Sola, Editor (World Scientific, 1997);    
J.~Sola, Pramana {\bf 51}, 239 (1998); W. Hollik and C. Schappacher,    
Nucl. Phys. {\bf B545}, 98 (1999);    
D.~Garcia and J.~Sola,    
Phys. Lett. {\bf B357}, 349 (1995); D.~Garcia, J.~Jiminez and J.~Sola,    
Phys. Lett. {\bf B347}, 309 (1995) and {\bf 347}, 321 (1995).    
    
%
\bibitem{zerwas} Some representative examples include, M.~D\'iaz and    
D.~Ross, hep-ph/0205257 (2002);    
J.~Gausch, W.~Hollik and J.~Sola, Phys. Lett. {\bf B510}, 211 (2001);    
W. Beenakker, T.~Blank and W.~Hollik, hep-ph/0011092 (2000);    
R. Hopker, M. Spira and P.M. Zerwas, Nucl. Phys. {\bf B492}, 51 (1997);      
W. Beenakker, R. Hopker, and P.M. Zerwas, Phys. Lett. {\bf B378}, 159      
(1996); E.~Berger, M.~Klasen and T.~Tait, Phys. Rev. {\bf D62}, 095014  
(2000); M.~Berger, B.~Harris, M.~Klasen and T.~Tait, hep-ph/9903237 (1999);  
W. Beenakker {\it et al.} Phys. Rev. Lett.  {\bf 83}, 3780 (1999);    
M.~A.~Diaz, S.~F.~King and D.~A.~Ross, Nucl.\ Phys.\ B {\bf 529}, 23 (1998);    
H.~Eberl, A.~Bartl and W.~Majerotto, Nucl.\ Phys.\ B {\bf 472}, 481 (1996).    
      
%
\bibitem{carena}R.~Hempfling, Phys. Rev. {\bf D49}, 6168 (1994);      
L.~Hall, R.~Rattazzi and U.~Sarid, Phys. Rev. {\bf D50},      
7048 (1994); M.~Carena {\it et al.} Nucl. Phys. {\bf B426}, 269      
(1994);.~Rattazzi and U.~Sarid, Phys. Rev. {\bf D53}, 1553 (1996).       
%
%
\bibitem{large} H.~Baer {\it et al.} Phys. Rev. Lett. {\bf 79}, 986 (1997);    
{\it ibid} {\bf 80}, 642 (1998) (E)    
(1997);   H.~Baer {\it et al.} Phys. Rev. {\bf D58}, 075008 (1998);    
H.~Baer {\it et al.} Phys. Rev. {\bf D59}, 055014 (1999).    
%
\bibitem{soten} V. Barger, M. Berger and P. Ohmann,       
Phys. Rev. D{\bf 49}, 4908 (1994); M.~Carena and C.~Wagner, CERN-TH-7321      
(1994); H.~Murayama, M.~Olechowski and S.~Pokorski, Phys. Lett. {\bf      
B371}, 57 (1996).       
%
\bibitem{superk} Y.~Fukuda {\it et al.} Phys. Rev. Lett. {\bf 82}, 2644      
(1999) and {\bf 85}, 3999 (2000).       
%
\bibitem{us} H.~Baer {\it et al.}, Phys. Rev. {\bf D61}, 111701 (2000);      
and {\bf D63}, 015007 (2000); H.~Baer and J.~Ferrandis,      
Phys. Rev. Lett. {\bf 87}, 211803 (2001).      
%
\bibitem{raby}        
T. Blazek, R. Dermisek and  S. Raby, Phys. Rev. Lett. {\bf 88}, 111804      
(2002) and hep-ph/0201081 (2002), and references therein.      
%
\bibitem{Siegel:1979wq}      
W.~Siegel,      
Phys.\ Lett.\ B {\bf 84}, 193 (1979).      


\bibitem{mass} 
K. Melnikov and A. Yelkhovsky, Phys. Rev. {\bf D59}, 114009 (1999);
A.H. Hoang, Phys. Rev. {\bf D61}, 034005 (2000); 
M. Beneke and A. Signer, Phys. Lett. {\bf B471}, 233 (1999);
A.~A.~Penin and A.~A.~Pivovarov, 
Nucl.\ Phys. {\bf B549}, 217 (1999).

\bibitem{Groom:in}      
D.~E.~Groom {\it et al.}  [Particle Data Group Collaboration],      
Eur.\ Phys.\ J.\  {\bf C15} (2000) 1.      
%
%
\bibitem{note} Here, and in the remainder of the text, the function
$m_b^{\overline{\rm MS}}$ is evaluated at the scale equal to the running
mass; {\it i.e.} we drop the superscript $\overline{\rm MS}$ on the argument
of the function.

\bibitem{aronson} See, for instance, H.~Arason {\it et al.}
Phys. Rev. {\bf D46}, 3945 (1992); D.~Pierce {\it et al.},
Ref.~\cite{pierce}; L.~Hall {\it et al.}, Ref.~\cite{carena}.

\bibitem{shifman} {\it At the Frontier of Particle Physics: Handbook of    
QCD}, M.~Shifman, Editor (World Scientific, 2001).     
      
\bibitem{Tarrach:1980up}      
R.~Tarrach,      
Nucl.\ Phys.  {\bf B183}, 384 (1981).      

\bibitem{Gray:1990yh}      
N.~Gray, D.~J.~Broadhurst, W.~Grafe and K.~Schilcher,      
Z.\ Phys. {\bf C48}, 673 (1990).      
    
\bibitem{Fleischer:1998dw}      
J.~Fleischer, F.~Jegerlehner, O.~V.~Tarasov and O.~L.~Veretin,      
Nucl.\ Phys. {\bf b539}, 671 (1999)      
[Erratum-ibid. {\bf B571}, 511]      
    
      
      
      
\bibitem{Chetyrkin:1999qi}      
K.~G.~Chetyrkin and M.~Steinhauser,      
Nucl.\ Phys. {\bf B573}, 617 (2000)      
      
      
\bibitem{Chetyrkin:2000yt}      
K.~G.~Chetyrkin, J.~H.~Kuhn and M.~Steinhauser,      
Comp. Phys. Comm. {\bf 133}, 43 (2000)      
    
    
    
\bibitem{Melnikov:2000qh}      
K.~Melnikov and T.~v.~Ritbergen,      
Phys.\ Lett. {\bf B482}, 99 (2000)      
      
\bibitem{Avdeev:1997sz}      
L.~V.~Avdeev and M.~Y.~Kalmykov,      
Nucl.\ Phys. {\bf B502}, 419 (1997)      
%

      
\bibitem{Tarasov:au}      
O.~V.~Tarasov, A.~A.~Vladimirov and A.~Y.~Zharkov,      
Phys.\ Lett.\ B {\bf 93} (1980) 429      
      
\bibitem{Chetyrkin:1997dh}      
K.~G.~Chetyrkin,      
Phys.\ Lett.\ B {\bf 404}, 161 (1997)      
    
      
\bibitem{Vermaseren:1997fq}      
J.~A.~Vermaseren, S.~A.~Larin and T.~van Ritbergen,      
Phys.\ Lett.\ {\bf B405}, 327 (1997)      
    
      
      
\bibitem{Chetyrkin:1997sg}      
K.~G.~Chetyrkin, B.~A.~Kniehl and M.~Steinhauser,      
Phys.\ Rev.\ Lett.\  {\bf 79}, 2184 (1997)      
      
\bibitem{Antoniadis:1982vr}      
I.~Antoniadis, C.~Kounnas and K.~Tamvakis,      
Phys.\ Lett.\ B {\bf 119}, 377 (1982).      
%
\bibitem{note2} Analogous reasoning also provides an understanding why 
$\log({m \over \mu})$ terms appear in the relations between the pole
mass and
short distance masses.
%
\bibitem{delphi} P.~Abreu {\it et al.},
Phys. Lett. {\bf B418}, 430 (1998).    



\end{references}
\end{document}